\documentclass[conference]{IEEEtran}
\usepackage{amsmath,amsthm,amsfonts,amssymb,bm,mathrsfs}
\usepackage{graphicx,subfigure}
\usepackage{cite}
\usepackage{mathrsfs}
\usepackage{algorithm}
\usepackage{algorithmic}
\usepackage{multirow,booktabs,threeparttable}
\usepackage{tablefootnote}
\usepackage{color}
\usepackage{verbatim}
\graphicspath{{fig/}}

\theoremstyle{remark}

\theoremstyle{plain}
\newtheorem*{remark}{Remark}
\begin{document}
\title{Traffic Prediction Based on Random Connectivity in Deep Learning with Long Short-Term Memory}


\author{\IEEEauthorblockN{Yuxiu Hua\IEEEauthorrefmark{1}, Zhifeng 		Zhao\IEEEauthorrefmark{1}, Rongpeng Li\IEEEauthorrefmark{1}, Xianfu Chen\IEEEauthorrefmark{2}, Zhiming Liu\IEEEauthorrefmark{3}, Honggang Zhang\IEEEauthorrefmark{1}}
\IEEEauthorblockA{\IEEEauthorrefmark{1,2,3,6}College of Information Science and Electronic Engineering\\Zhejiang University, Zheda Road 38, Hangzhou 310027, China\\Emails:\{21631087, zhaozf, lirongpeng,  honggangzhang\}@zju.edu.cn}
\IEEEauthorblockA{\IEEEauthorrefmark{2}VTT Technical Research Centre of Finland, P.O. Box 1100, FI-90571 Oulu, Finland\\Email:Xianfu.Chen@vtt.fi}
\IEEEauthorblockA{\IEEEauthorrefmark{3}China Mobile Research Institute, Beijing 100053, China\\Email:liuzhiming@chinamobile.com}}


\maketitle
\begin{abstract}
Traffic prediction plays an important role in evaluating the performance of telecommunication networks and attracts intense research interests. A significant number of algorithms and models have been put forward to analyse traffic data and make prediction. In the recent big data era, deep learning has been exploited to 
mine the profound information hidden in the data. In particular, Long Short-Term Memory (LSTM), one kind of Recurrent Neural Network (RNN) schemes, has attracted a lot of attentions due to its capability of processing the long-range dependency embedded in the sequential traffic data. However, LSTM has considerable computational cost, which can not be tolerated in tasks with stringent latency requirement. In this paper, we propose a deep learning model based on LSTM, called Random Connectivity LSTM (RCLSTM). Compared to the conventional LSTM, RCLSTM makes a notable breakthrough in the formation of neural network, which is that the neurons are connected in a stochastic manner rather than full connected. So, the RCLSTM, with certain intrinsic sparsity, have many neural connections absent {\tiny} (distinguished from the full connectivity) and which leads to the reduction of the  parameters to be trained and the computational cost.
We apply the RCLSTM to predict traffic and validate that the RCLSTM with even 35\% neural connectivity still shows a satisfactory performance. When we gradually add training samples, the performance of RCLSTM becomes increasingly closer to the baseline LSTM. Moreover, for the input traffic sequences of enough length,
the RCLSTM exhibits even superior prediction accuracy than the baseline LSTM.

\end{abstract}

\section{Introduction}
With the proliferation of mobile terminals and the astonishing expansion of Mobile Internet, Internet of Things (IoT) and cloud computing, the mobile communication network has become an indispensable social infrastructure, which is bound up with people's lives and many social domains. Cisco's latest statistical result shows that mobile data traffic has grown 18-fold over the past 5 years and it will increase 7-fold between 2016 and 2021 \cite{cisco}. These trends inevitably lead to explosive growth in the size and complexity of communication networks, which leaves a series of challenging issues to be addressed. For example, if the running status of communication networks can be monitored and fine-tuned promptly or even in advance, the networks' stability and the user experience will be greatly improved. Since data traffic is an important dimension to measure the performance and the running status of networks, traffic prediction is of fundamental significance to the optimization and management of communication networks, such as optimal routing \& scheduling, energy saving, and network anomaly detection \cite{li2017learning}. Therefore, a precise traffic prediction takes an valuable role in planning, management and design of the communication networks.

Traffic prediction is a classical research field, which has attracted a lot of attentions and great efforts in developing various algorithms and protocols for wireless networks to utilize the resources efficiently and effectively \cite{wang2017spatiotemporal}.
Several methods for traffic prediction have been proposed in literature, which can be classified into two categories: linear methods and nonlinear methods \cite{DBLP:journals/corr/AzzouniP17}. For the aspect of linear prediction methods, the most widely adopted algorithms are based on AutoRegression Integrated Moving Average (ARIMA) \cite{Zhou2006}. However, ARIMA models have a severe limitation with their natural tendency to concentrate on the mean values of the past series data. Therefore, it is almost unable to capture the rapid variational process underlying the traffic load \cite{hong2012application}. On the other hand, the most commonly used nonlinear algorithm is Support Vector Regression (SVR) \cite{castro2009online}. The success of SVR lies in four factors, which include good generation, global optimization solution, the ability to handle nonlinear problem, and the sparseness of the solution \cite{alwee2013hybrid}. Just like a coin has pro and con, SVR is limited by the lack of structured means to determine some key parameters in the model, thus incurring the deficiency of knowledge on how to select the key parameters \cite{hong2012application}. 


In addition to the above conventional algorithms, more and more studies have adopted Artificial Neural Networks (ANNs) for traffic prediction in recent years\cite{wang2017spatiotemporal}, \cite{DBLP:journals/corr/AzzouniP17}, \cite{nikravesh2016mobile}. 
Nikravesh \emph{et al.} have investigated the accuracy of Multi-Layer Perceptron (MLP) (a typical architecture of ANNs) in predicting future behavior of mobile network traffic when compared to Support Vector Machine (SVM) and the results show MLP has better accuracy than SVM \cite{nikravesh2016mobile}. Wang \emph{et al.} used Local Stacked AutoEncoders (LSAEs) and Global Stacked AutoEncoder (GSAE) to extract local traffic characteristics and then utilized Long Short-Term Memory (LSTM) to predict the traffic of cellular networks \cite{wang2017spatiotemporal}. As a universal approximator that can efficiently approach a continuous function with a desired level of accuracy, ANNs are verified by their effectiveness in solving nonlinear problems as long as they contain a sufficient number of parameters. However, if the number of ANNs parameters become extremely large, the total amount of computation and training time will become too heavy to be applicable.

Inspired by the interesting finding that the Convolutional Neural Networks (CNNs) with sparse neural connections have a similar or even superior performance in many experiments when compared to the conventional CNNs \cite{shafiee2016stochasticnet}, we propose a novel model (i.e. Random Connectivity LSTM, RCLSTM) based on LSTM structure, which contains fewer parameters than a conventional LSTM. Moreover, in this paper, 
we construct a three-layer stack RCLSTM network\footnote{Hereinafter, if not specifically mentioned, the ``network'' means ``neural network'', rather than ``mobile network'' or ``telecommunications network''.} to complete the task of traffic prediction. The experiment results exhibit that the RCLSTM network achieves a competitive performance when compared to the conventional LSTM network, and specifically the RCLSTM network with 35\% neural connectivity still possesses a strong capability in traffic prediction, which validates the effectiveness of the RCLSTM approach. What is more interesting is that when the length of input traffic sequences increases, the RCLSTM outperforms the conventional LSTM in terms of the prediction accuracy. To the best of our knowledge, we are the first to leverage the random connectivity for the formation of neural connections in LSTM memory block, and investigate the effectiveness and superiority of the RCLSTM with real traffic data.

The rest of this paper is organized as following. Section 2 presents the LSTM architecture and builds the RCLSTM model. Experiments design and results analyses are given in Section 3. Finally, Section 4 gives conclusions for this paper and points out the future work.

\section{Background and RCLSTM}
\label{sec:model construction based LSTM}
\subsection{Mathematical Background}
\label{sec:LSTM structure}
Basically, as illustrated in Fig. 1, ANNs can be classified into two kind of paradigms. 
The first is Feed Forward Neural Networks (FFNNs), which has no cycles formed in neural connections and is widely used in many fields such as data classification, object recognition, imaging processing, etc. 
FFNNs operate on fixed-size window of input data. So, they can only model the data within the window and are unsuitable for handling historical dependencies. They can provide limited capability of temporal modeling and traffic prediction. By contrast, different from FFNNs, the recurrent connections in Recurrent Neural Networks (RNNs) allow the information of historical inputs to be stored in the network's internal state, and thereby make them capable of mapping the whole historical input data to each output in principle. In theory, RNN can learn the knowledge of temporal sequence without length constraints. However, for any standard RNN architecture, the influence of a given input on the hidden layers and finally on the network output, would either decay or blow up exponentially when cycling around the network's recurrent connections. To address these problems, an elegant RNN architecture (i.e., LSTM) has been designed \cite{hochreiter1997long}. LSTM has shown cutting-edge performance for language modeling, handwriting recognition \cite{liwicki2007novel}, and phonetic labeling of acoustic frames \cite{sak2014long}. 

\begin{figure}
	\begin{center}
		\includegraphics[trim=5mm 0mm 5mm 5mm,width=0.45\textwidth]{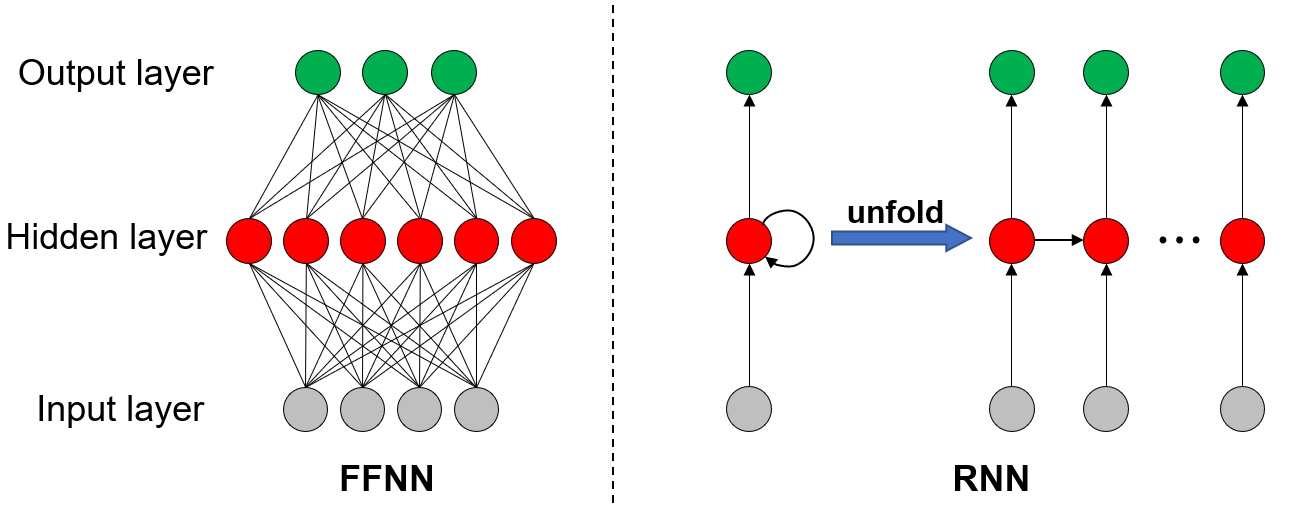}
		\setlength\abovecaptionskip{0pt}
		\setlength\belowcaptionskip{-5pt}
		\caption{Comparison of FFNN and RNN.}
		\label{fig1:lstm}
	\end{center}
\end{figure}

The key point that makes LSTM possess the ability to model long-term dependencies is a component called memory block. As illustrated in Fig. 2, memory block is a recurrently connected subnet, which contains some functional modules called memory cell and gates. The memory cell is in charge of remembering the temporal state of the network, while the gates composed of the sigmoid layers are responsible for controlling the amount of information flow. According to the corresponding practical functionalities, these gates are classified as input gate, output gate and forget gate. Input gate controls how much new information flows into the memory cell, while forget gate controls how much information still remains in the current memory cell through recurrent connection, and output gate controls how much information is used to compute the output activation of the memory block and further flows into the rest of the network.

\begin{figure}
	\begin{center}
		\includegraphics[trim=5mm 0mm 5mm 5mm,width=0.45\textwidth]{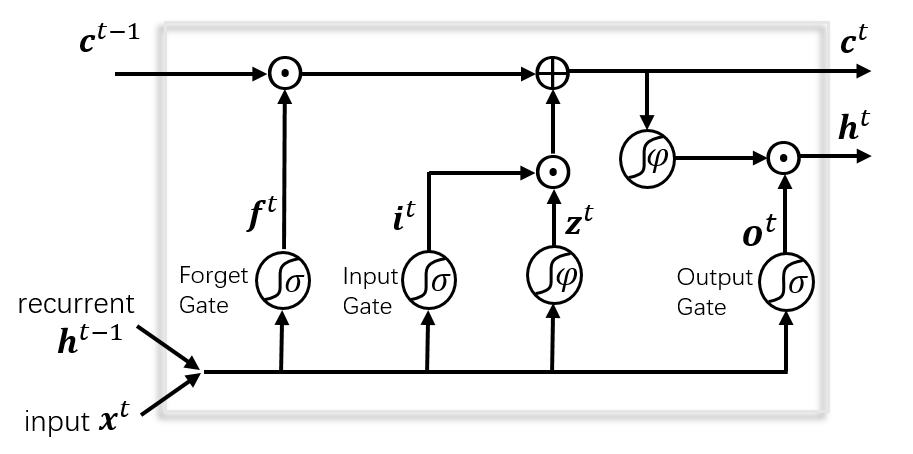}
		\setlength\abovecaptionskip{0pt}
		\setlength\belowcaptionskip{-5pt}
		\caption{An illustration of LSTM memory block.}
		\label{fig1:lstm}
	\end{center}
\end{figure}

Suppose that the input vector at time $t$ is $\textbf{x}^t$, the hidden state vector at time $t$ is $\textbf{h}^t$, the memory cell at time $t$ is $\textbf{c}^t$. The $\textbf{i}$, $\textbf{o}$ and $\textbf{f}$ denote the output of input gate, output gate and forget gate, respectively; and $\textbf{z}$ denotes input activation. $\odot$ and $\oplus$ denote dot product and summation of two vectors. $\sigma$$(\cdot)$ usually takes the sigmoid function, and $\varphi$$(\cdot)$ is usually the hyperbolic tangent function\footnote{Interested readers could refer to \cite{goodfellow2016deep} (and references therein) for more details.}, respectively:

\begin{align}
\setlength{\abovedisplayskip}{1pt}
\setlength{\belowdisplayskip}{1pt}
\sigma(\emph{x}) &= \frac{1}{1+\emph{e}^{-\emph{x}}} \\
\varphi(\emph{x}) &= 2\sigma(2\emph{x})-1
\end{align}
\\*
After these definitions, whole functionalities of the LSTM memory block implementation are formulated as follows:
 
\begin{align}
\setlength{\abovedisplayskip}{1pt}
\setlength{\belowdisplayskip}{1pt}
\textbf{i}^t &= \sigma(\textbf{W}_{xi}\textbf{x}^t + \textbf{W}_{hi}\textbf{h}^{t-1} + \textbf{b}_i) \\
\textbf{f}^t &= \sigma(\textbf{W}_{xf}\textbf{x}^t + \textbf{W}_{hf}\textbf{h}^{t-1} + \textbf{b}_f) \\
\textbf{o}^t &= \sigma(\textbf{W}_{xo}\textbf{x}^t + \textbf{W}_{ho}\textbf{h}^{t-1} + \textbf{b}_o) \\
\textbf{z}^t &= \varphi(\textbf{W}_{xc}\textbf{x}^t + \textbf{W}_{hc}\textbf{h}^{t-1} + \textbf{b}_c) \\
\textbf{c}^t &= \textbf{f}^t\odot\textbf{c}^{t-1} + \textbf{i}^t\odot\textbf{z}^t \\
\textbf{h}^t &= \textbf{o}^t\odot\varphi(\textbf{c}^t)
\end{align}
\\*
where the $\textbf{W}$ terms with different subscripts denote different weight matrices (e.g., $\textbf{W}_{hf}$ is the hidden-forget weight matrix), and similarly the $\textbf{b}$ terms with different subscripts denote different bias vectors (e.g., $\textbf{b}_f$ is the bias vector for forget gate).

\subsection{Random Connectivity LSTM}
\label{RCLSTM}
In fact, the conventional LSTM (including its variants) basically follows the classical pattern that the neural connections in memory block all exist and one cannot change the neural connectivity in networks. However, neuroscience researchers have found that for specific functional connectivity in neural microcircuits, random topology formation of synapses can provide a sufficient foundation \cite{hill2012statistical}. This discovery stands on the opposite side of the conventional cases, where that neural connectivity is considered to be more heuristic so that neurons need to be connected in a more fully organized manner. In regard to LSTM, it does beg the question as to whether a formation strategy of more random neural connectivity like in the brain may yield potential benefits to LSTM performance and efficiency. Holding with this idea, we build up the Random Connectivity LSTM (RCLSTM) model.

For simplicity of representation, we reformulate the equations from Eq. (3) to Eq. (8). Firstly, we concatenate $\textbf{x}^t$ and $\textbf{h}^{t-1}$, and this operation is denoted as $[\textbf{x}^t,\textbf{h}^{t-1}]$. Then, we regard $\textbf{W}$ terms as the weight matrices between $[\textbf{x}^t,\textbf{h}^{t-1}]$ and the gates (e.g., $\textbf{W}_f$ denotes the forget weight matrix). After these definitions, the reformulated equations are obtained as: 

\begin{align}
\setlength{\abovedisplayskip}{1pt}
\setlength{\belowdisplayskip}{1pt}
\textbf{i}^t &= \sigma(\textbf{W}_i[\textbf{x}^t,\textbf{h}^{t-1}] + \textbf{b}_i) \\
\textbf{f}^t &= \sigma(\textbf{W}_f[\textbf{x}^t,\textbf{h}^{t-1}] + \textbf{b}_f) \\
\textbf{o}^t &= \sigma(\textbf{W}_o[\textbf{x}^t,\textbf{h}^{t-1}] + \textbf{b}_o) \\
\textbf{z}^t &= \varphi(\textbf{W}_c[\textbf{x}^t,\textbf{h}^{t-1}] + \textbf{b}_c) \\
\textbf{c}^t &= \textbf{f}^t\odot\textbf{c}^{t-1} + \textbf{i}^t\odot\textbf{z}^t \\
\textbf{h}^t &= \textbf{o}^t\odot\varphi(\textbf{c}^t)
\end{align}
\\*
It is easy to prove that the above formulas are equivalent to the corresponding equations Eq. (3) to Eq. (8).


Actually, the parameters that require to be trained in LSTM only exist between the input part (i.e., $[\textbf{x}^t,\textbf{h}^{t-1}]$) and the functional layers (i.e., the gates layers and the tanh layer for input activation). Hence, if we regard the input part as the input layer and the conjunction of functional layers as the output layer, then a LSTM memory block can be transformed to a simple two-layer FFNN.
Accordingly, LSTM memory block can be represented as a random graph $\mathcal{G}(\mathcal{V}, \emph{p})$, where $\mathcal{V}$ denotes the set of neurons $\mathcal{V}=\{\emph{v}_{lk}| l\subset{\{1,2\}}, 1\leqslant k\leqslant \emph{n}_l\}$, with $\emph{v}_{lk}$ representing the $\emph{k}^{th}$ neuron at layer $l$, $\emph{n}_l$ representing the number of neurons at layer $l$, and $\emph{p}$ representing the probabilities that neural connections occur between neurons (e.g., $\emph{p}[i\rightarrow{j}]$ represents the probability that a neural connection occurs between the neurons $\emph{v}_{1i}$ and $\emph{v}_{2j}$).


\begin{figure*}
	\begin{center}
		\includegraphics[trim=5mm 0mm 5mm 5mm,width=0.85\textwidth]{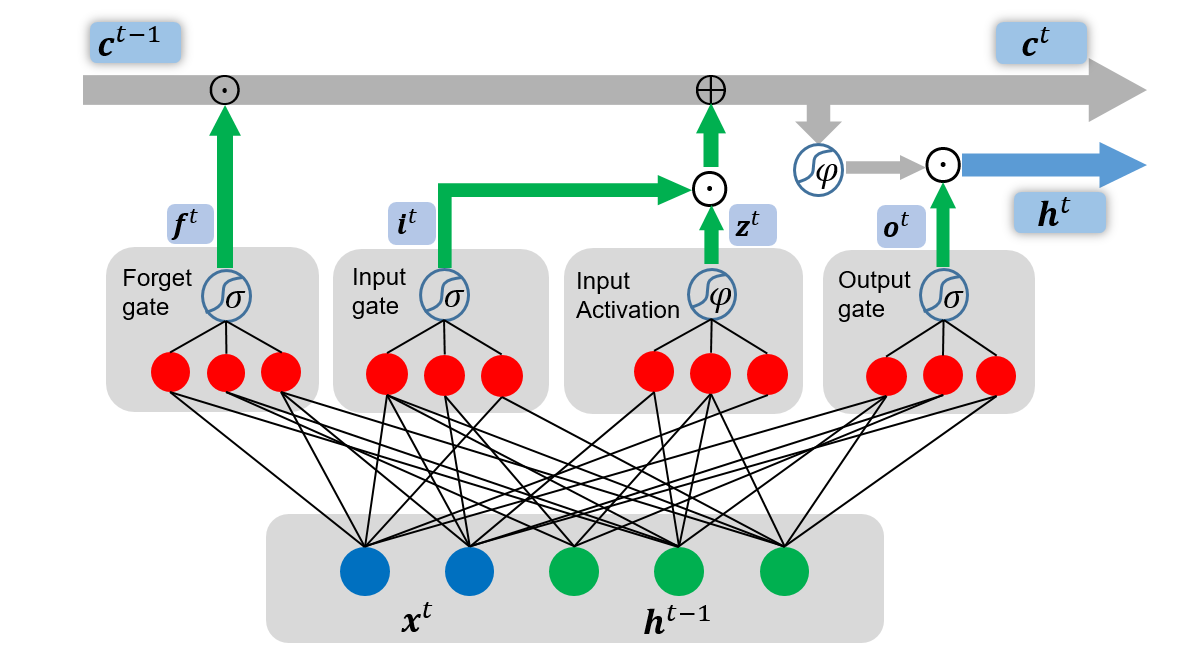}
		\setlength\abovecaptionskip{0pt}
		\setlength\belowcaptionskip{-5pt}
		\caption{A representative example of RCLSTM block.}
		\label{fig3:example}
	\end{center}
\end{figure*}

Using the definitions above, one can reform the neural connections within a LSTM memory block as a realization of the random graph 
$\mathcal{G}(\mathcal{V}, \emph{p})$ by initializing with a set of neurons $\mathcal{V}$, meanwhile randomly inserting neural connections among the set of neurons independently in a stochastic manner with probabilities of 
$\emph{p}$. The strategy to establish neural connections in RCLSTM is:

{\setlength\abovedisplayskip{1pt}
\setlength\belowdisplayskip{1pt}
\begin{equation}
\emph{e}[i\rightarrow{j}] \ \text{exists} \ \text{where} \ \emph{p}[i\rightarrow{j}]\geqslant{T}
\end{equation}}
\\*
where $\emph{e}[i\rightarrow{j}]$ denotes the neural connection between neuron $\emph{v}_{1i}$ and neuron $\emph{v}_{2j}$, and $T$ is a threshold which indicates the sparsity of neural connectivity in RCLSTM. 

Based on the aforementioned model, we construct a representative RCLSTM as illustrated in Fig. 3. It can be observed that the RCLSTM has a large number of distinct formation because of the random nature of the formation process during neural connections. Accordingly, by randomly forming the structure of LSTM memory block, the corresponding neural network can be highly sparse and bring considerable decrease in the number of involved parameters as well as the computational loads.

\begin{remark}
The RCLSTM manifests a strong capability in traffic prediction while the number of parameters to be trained is reduced, which in effect decreases the computational loads and complexity. Moreover, the RCLSTM exhibits superior performance than the conventional LSTM when the length of input traffic sequences increases.
\end{remark}


\vspace{-5pt}
\section{ Numerical Experiments}
\label{sec:experiment}
In this section, our goal is to verify the performance of RCLSTM in traffic prediction. In order to achieve this purpose, we construct a three-layer RNN network with the newly designed RCLSTM memory block. Its structure and unfolding form are described in Fig. 4. Firstly, we take advantage of the model to predict traffic, and explore the changing features of prediction accuracy when using RCLSTM memory blocks with different sparsity patterns generated from the random neural connections. Then, we gradually reduce the number of training data samples to investigate the effect of the size of training data set on the prediction accuracy. Finally, we use different length of input traffic sequences to train the RCLSTM network and explore the capability of the RCLSTM to model long-term dependencies compared to the conventional LSTM. Recently, Godfrey\emph{et al.} \cite{DBLP:journals/corr/GodfreyG17} have proved that LSTM performs better than ARIMA and SVR on time-series forecasting through a large number of experiments. Therefore, we only discuss the comparison between the RCLSTM and the conventional LSTM.

\begin{figure}
	\includegraphics[trim=5mm 0mm 5mm 5mm,width=0.45\textwidth]{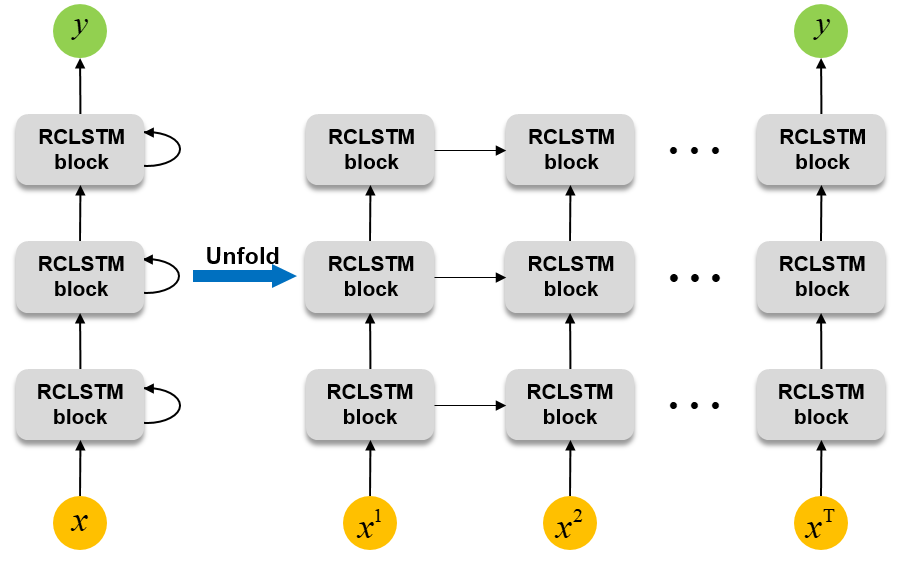}
	\setlength\abovecaptionskip{0pt}
	\setlength\belowcaptionskip{-5pt}
	\caption{The designed RCLSTM network for numerical experiment.}
	\label{fig4：experiment}
\end{figure}

\vspace{-6pt}
\subsection{Traffic Data Description and Processing}
Here we will evaluate the model's performance on traffic prediction depending upon real traffic data from the G\'EANT backbone networks\footnote{https://www.geant.org/Projects/GEANT\_Project\_GN4}. G\'EANT is a pan-European data source for the research and education communities. It builds up the European National Research and Education Networks (NRENs) which interconnect various universities and research institutions. The traffic data are obtained from the G\'EANT network by 15-min interval for several months. In this study, because of the absence of some data, we only select 7289 traffic data from 76 workdays measured between 2005-01-01 00:00AM and 2005-04-30 00:00AM. Then we normalize these data samples according to the following formula:

\begin{equation}
\setlength{\abovedisplayskip}{1pt}
\setlength{\belowdisplayskip}{1pt}
\dfrac{\emph{\textbf{x}}-\mu(\emph{\textbf{x}})}{\sigma(\emph{\textbf{x}})}
\end{equation}
\\*
where $\emph{\textbf{x}}$ is the original data, $\mu(\emph{\textbf{x}})$ denotes the average of $\emph{\textbf{x}}$, and $\sigma(\emph{\textbf{x}})$ denotes the standard deviation of $\emph{\textbf{x}}$. Because the number of the traffic data we obtained is so large that we cannot depict them neatly in a graph. Therefore, Fig. 5 describes some intercepted traffic data after normalization. Next, we apply a variable sliding window to specify a fixed number of previous timeslots to learn before predicting the current traffic data. This operation is to deal with the issue that the total number of timeslots may be too big to compute. Finally, we split the processed data into two sets (i.e., training set and test set). The training set is used to train the RCLSTM network while the test set is used to evaluate and validate its prediction accuracy.

\begin{figure}
	\begin{center}
		\includegraphics[trim=5mm 0mm 5mm 5mm,width=0.45\textwidth]{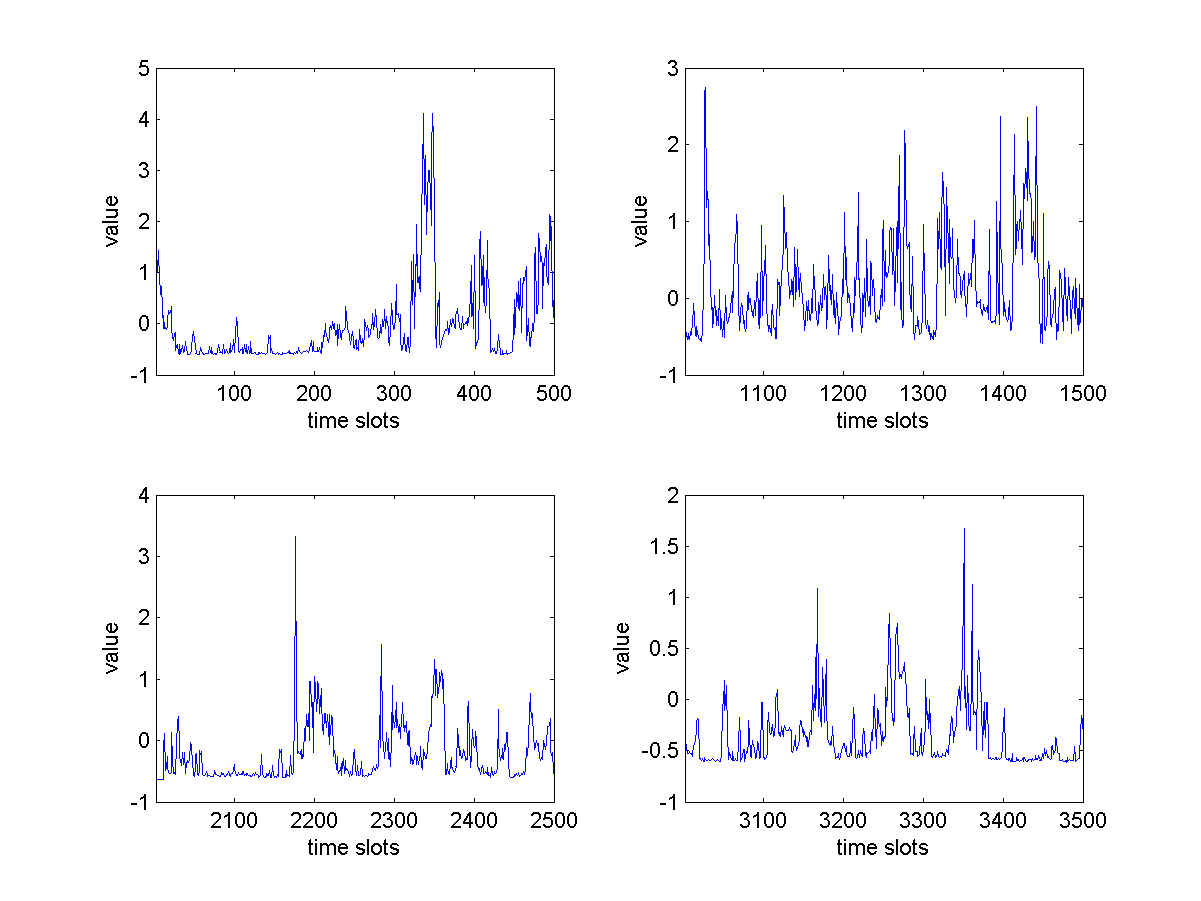}
		\setlength\abovecaptionskip{0pt}
		\setlength\belowcaptionskip{-5pt}
		\caption{The intercepted traffic data after normalization.}
		\label{fig5：mse_mae}
	\end{center}
\end{figure}

\subsection{Evaluation Metrics}
Evaluation metrics are necessary to evaluate the performance of our traffic prediction model. Accordingly, Mean Square Error (MSE) and Mean Absolute Error (MAE) are used to estimate the prediction accuracy. MSE measures the average of squared errors, which quantifies the difference between the predicted values and the actual values. MAE, defined as the average of absolute error, is a measure of difference between two variables as well. The expressions of MSE and MAE are mathematically formulated as:

\begin{align}
\setlength{\abovedisplayskip}{1pt}
\setlength{\belowdisplayskip}{1pt}
MSE &= \frac{1}{N}\sum_{i=1}^{N}(y_i-\hat{y_i})^2 \\
MAE &= \frac{1}{N}\sum_{i=1}^{N}\left|y_i-\hat{y_i}\right|
\end{align} 
\\*
where $\emph{y}_i$ is the actual value, $\hat{\emph{y}_i}$ is the predicted value and $N$ represents the total number of predictions.

\subsection{Testing Results and Analyses}
\begin{figure}
	\begin{center}
		
		\includegraphics[trim=5mm 0mm 5mm 5mm,width=0.45\textwidth]{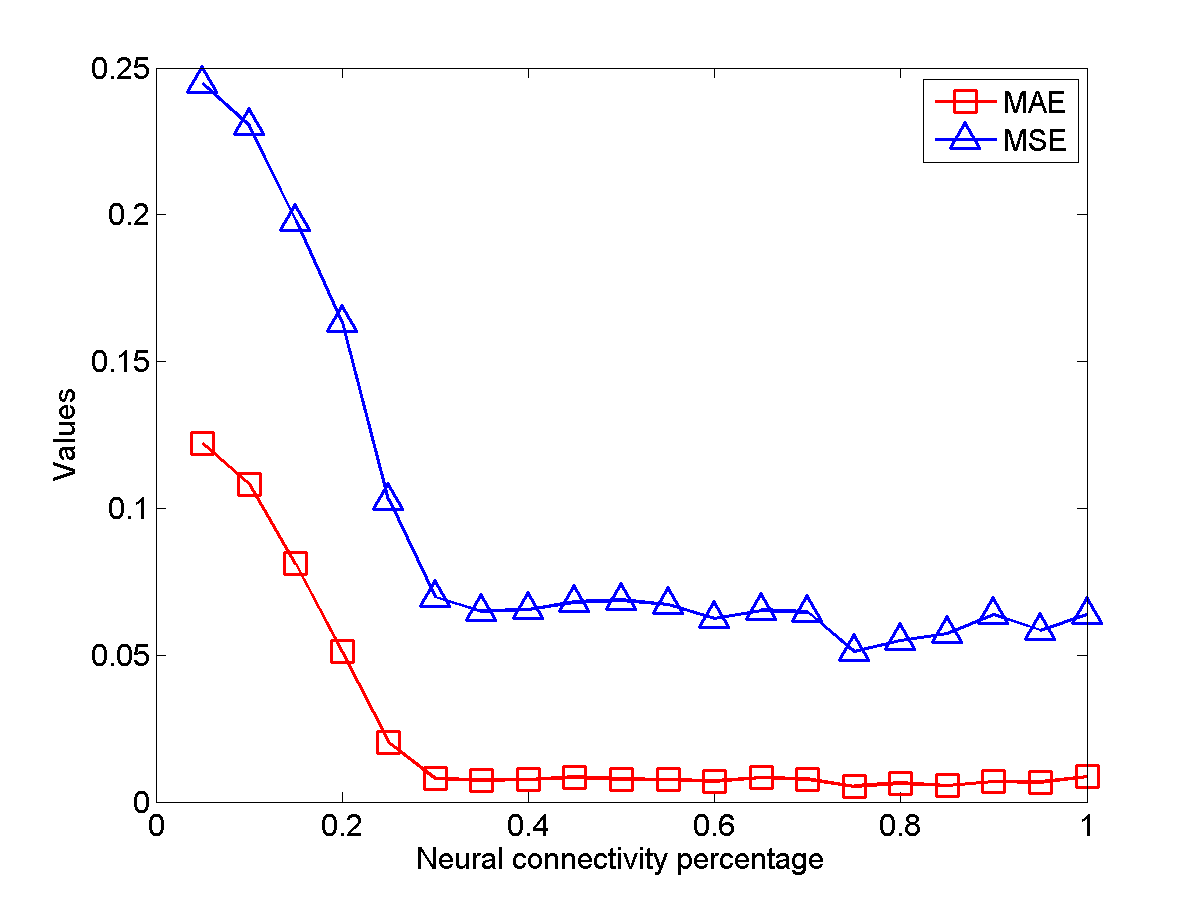}
		\setlength\abovecaptionskip{0pt}
		\setlength\belowcaptionskip{-5pt}
		\caption{MSE and MAE over the percentage of neural connectivity.}
		\label{fig5：mse_mae}
	\end{center}
\end{figure}

\begin{figure}
	\begin{center}
		\includegraphics[trim=5mm 0mm 5mm 5mm,width=0.45\textwidth]{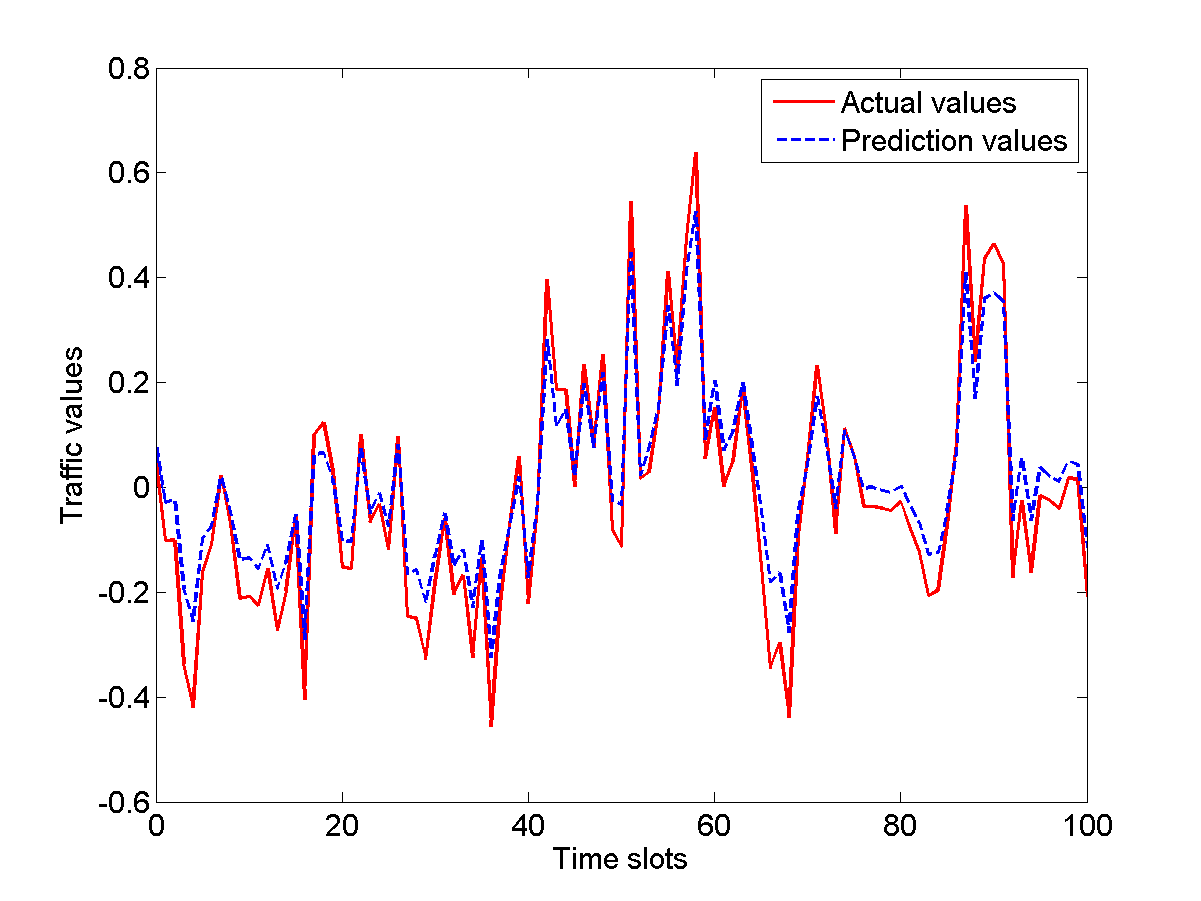}
		\setlength\abovecaptionskip{0pt}
		\setlength\belowcaptionskip{-5pt}
		\caption{Comparison of the actual and predicted traffic.}
	\end{center}
\end{figure}


Fig. 6 shows the MSE and MAE under different percentages of neural connectivity in RCLSTM (note that 100\% connectivity means the baseline LSTM), where the connection probability obeys a uniform distribution ($\emph{p}_g\sim U(0,1)$). It can be observed from Fig. 6 that neural connectivity of 35\% is a turning point, below which the MSE and MAE are both serious (increasing rapidly to larger values) while above which the prediction accuracy stability remains with impressive performance. Fig. 7 provides the comparison of the actual and predicted traffic values when we set 35\% neural connectivity in RCLSTM. We observe that the predicted values can match the varying trend and features of the actual values very well. This means that the RCLSTM can attain strong predication ability while decreasing the number of neural connections, which in effect decreases the computational loads and complexity. 

\begin{figure}
	\begin{center}
		\includegraphics[trim=5mm 0mm 5mm 5mm,width=0.45\textwidth]{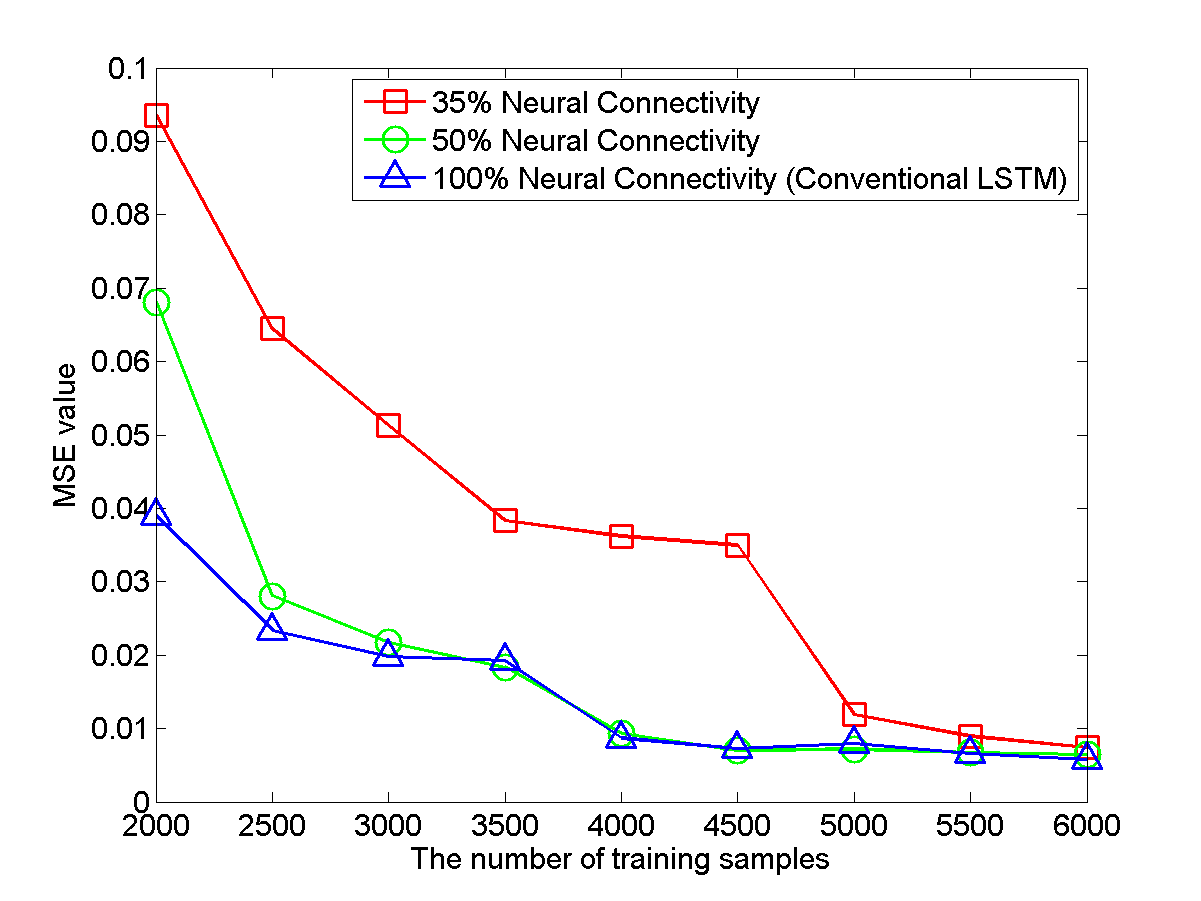}
		\setlength\abovecaptionskip{0pt}
		\setlength\belowcaptionskip{-5pt}
		\caption{MSE over the number of training samples.}
	\end{center}
\end{figure}

\begin{figure}
	\begin{center}
		\includegraphics[trim=5mm 0mm 5mm 5mm,width=0.45\textwidth]{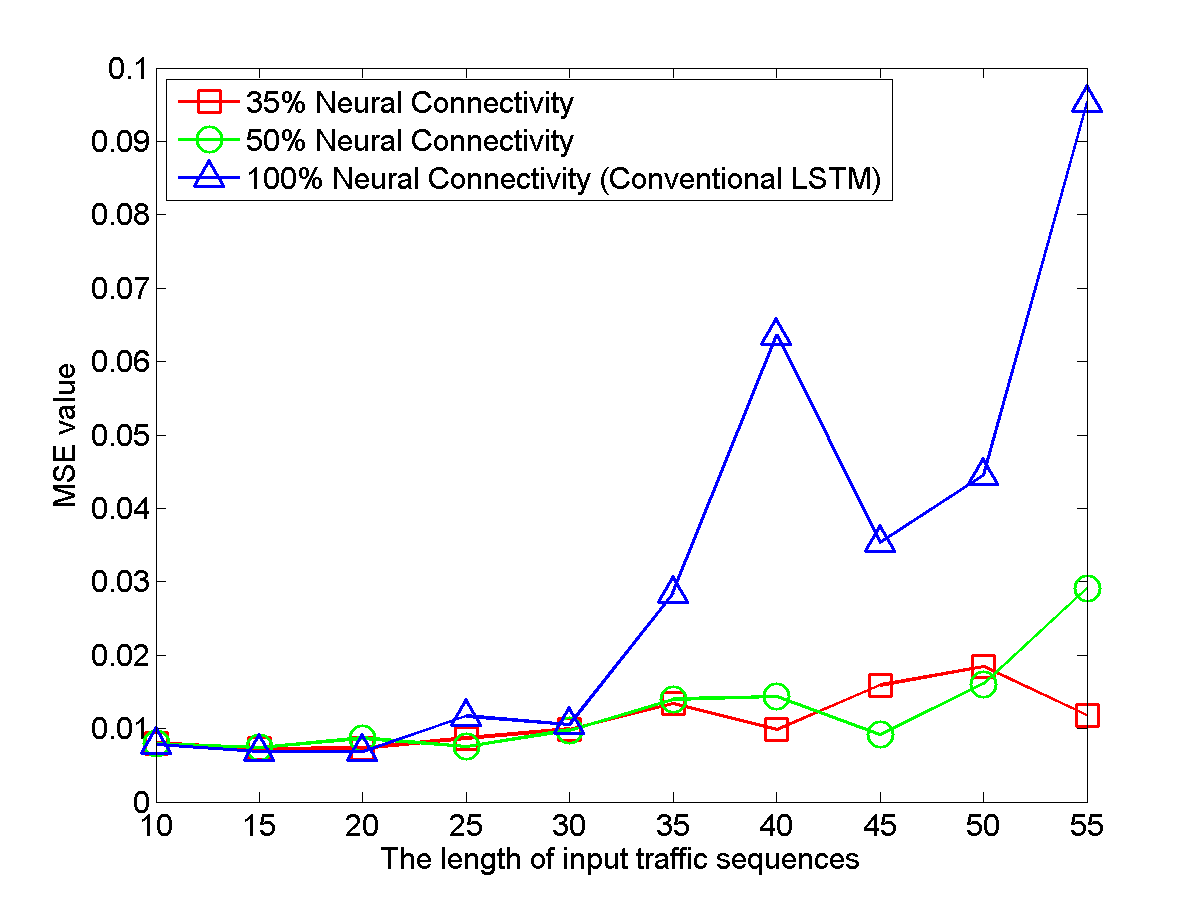}
		\setlength\abovecaptionskip{0pt}
		\setlength\belowcaptionskip{-5pt}
		\caption{MSE over the length of input traffic sequences.}
	\end{center}
\end{figure}

Then, we examine the prediction performance when the number of training samples varies in Fig. 8. When the number of training samples increases, the MSEs of both RCLSTMs and the conventional LSTM decrease. Moreover, when the number of training samples is less than 5000, there exists a big gap between the RCLSTM with 35\% neural connectivity and the other two settings. But the gap gradually disappears when the number of training samples increases. On the other hand, the RCLSTM with just 50\% neural connectivity is capable of achieving the performance similar to the conventional LSTM with full neural connections.

Furthermore, we investigate the capability of RCLSTM to characterize long-term dependencies by conducting a set of experiments based on different length of input traffic sequences obtained by variable sliding window, and the results are shown in Fig. 9. It can be observed that when the length of input traffic sequences increases, the MSE of the conventional LSTM varies greatly, while the MSEs of the RCLSTMs fluctuate faintly and become substantially lower than that of the conventional LSTM. These results indicate the interesting finding that RCLSTM has a better performance than the conventional LSTM when the input traffic sequences have sufficiently long timeslots.

\section{Conclusions}
In this paper, we have reinvestigated the field of traffic prediction with deep learning and proposed a new model named Random Connectivity LSTM (RCLSTM) based on the conventional LSTM. The key point of RCLSTM is to construct the neural network by forming and realizing a random graph and then grant a certain probability to the neural connectivity. We have checked the effectiveness of the RCLSTM model by applying it to predict actual traffic data, and validated that although the neural connections in the RCLSTM model are sparse, its performance is as satisfactory as the conventional LSTM network. Moreover, when the input data is a time series of considerable length, the RCLSTM is even better than the conventional LSTM. The RCLSTM can be used in a number of learning and prediction scenarios, but we need to further verify its contribution to decrease the computational complexity in a quantification manner, which will be our research works in the future.




\bibliographystyle{IEEEtran}
\bibliography{icc}

\end{document}